\documentstyle[12pt,twoside,psfig,fleqn,espcrc1]{article}

\newcommand{\AmS}{{\protect\the\textfont2
  A\kern-.1667em\lower.5ex\hbox{M}\kern-.125emS}}

\title{Bremsstrahlung of 350--450 MeV protons as a tool to study
                 $NN$ interaction off-shell\thanks{Talk given by Andrey
                 Shirokov at the International Conference on Quark Lepton
                 Nuclear Physics ``QULEN97'',
                 May 20-23, 1997, Osaka, Japan.}}

\author{   N.A. Khokhlov\address{Khabarovsk State Technical University,
 Khabarovsk 680035, Russia},
V.A. Knyr$^{\rm a}$,
V.G. Neudatchin$^{\rm b}$
and
Andrey M. Shirokov\address{Institute for Nuclear Physics, Moscow State University,
Moscow 119899, Russia}\thanks{E-mail: shirokov@anna19.npi.msu.su}  }

\begin{document}
\maketitle

\begin{abstract}
The $pp\to pp\gamma$ bremsstrahlung cross section is calculated within the
method of
coordinate space representation. It is shown that in the beam energy range
of 350--450~MeV a deep attractive
$NN$-potential with forbidden states (Moscow potential) and realistic meson
exchange potentials (MEP) give rise to the cross sections that differ
essentially in shape: the cross sections nearly coincide in the minima but
differ  by a factor of 5 approximately in the maxima. Therefore,
the $pp\to pp\gamma$ reaction at energies $\sim$350--450~Mev can be used to
study $NN$ interaction off-shell and to discriminate experimentally between
MEP and Moscow potential.
\end{abstract}

\bigskip
 The most popular of the $NN$ interactions today are meson exchange
potentials (MEP). The central component of the $NN$
interaction within the MEP model has a short-range repulsive core of the
radius $r_{\rm c} \simeq$~0.5--0.6~fm and is attractive at larger
distances. Modern ME potentials (Reid, Bonn, Paris, Argonne,
Nijmegen, etc.) are carefully fitted to the existing deuteron
and $NN$ scattering data up to the laboratory energies of $\sim$500~MeV
and higher. However, these data involve only the on-shell properties of the
interaction. The off-shell properties of the interaction are prominent in
the binding energies of few-body systems.  It is well-known that ME
potentials underbind the trinucleon and light nuclei, and this  problem
cannot be solved within the MEP model by allowing for realistic three-body
forces \cite{Pickl}.

An alternative to the MEP model is the model of a deep attractive $NN$
potential (the so-called Moscow potential, MP) \cite{Moscow1,MoscowProgr}.
MP supports additional deeply-bound states in $S$ and $P$ waves which are
treated as forbidden states.  The first excited state reproduces the
deuteron properties, the scattering observables are also well-described by
the latest version of MP \cite{MoscowProgr}. 
At short distances
where the MEP model wave function is suppressed by the repulsive core, the
MP wave function has an additional node in $S$ and $P$ waves that may be
interpreted as a manifestation of the 6-quark configurations
$s^4p^2[42]_{X}[42]_{CS}$ and $s^3p^3[33]_{X}[51]_{CS}$ in $S$ and $P$
waves, respectively.

MP and ME potentials are nearly equivalent on-shell at low and moderate
energies but differ essentially off-shell. However,
both MEP and MP underbind $^3$H. So, neither ME
nor Moscow potential model is favored by the few-body
verification of the $NN$ interaction off-shell. MEP model is based on the
well-developed ideas of meson exchange. At the same time, recent microscopic
studies of $NN$ interaction in a chiral constituent quark model
\cite{Gloz-NN} support the idea of the existence of the node of the $NN$
wave function at short distances. However, the amplitude of the oscillation
of the wave function at short range is suppressed as compared with the MP
wave function. So, MP and MEP can be treated as modern limiting
models of the interaction of real nucleons built of quarks.

Theoretical studies of $pp\to pp\gamma$ reaction  have been performed in a
number of papers (see, e.g.,
\cite{expt3,Brown,Heller,Fearing,39,41,25,deJong,Eden,vonGeramb}
and references therein). As it is concluded in a recent paper of J\"ade et
al~\cite{vonGeramb}, the result of numerous calculations of various authors
is very discouraging from the point of view of using $pp$ bremsstrahlung
as a comparative test of $NN$ potentials: the difference in the
$pp\to pp\gamma$ cross sections calculated with different $NN$ potentials
is too small to be measured experimentally. There are two reasons for this
disappointing result. First, as a rule only $NN$ potentials based on the
ideas of ME have been used in the calculations, and the $pp\to pp\gamma$
results are just the manifestation of the fact that there is nearly no
difference between various ME potentials. Next, usually the theoretical
investigations have been restricted to the beam energies not
exceeding 280~MeV that have been studied experimentally \cite{expt3}.
However, as we shall show below, the $pp\to pp\gamma$
reaction at energies 350--450~MeV can be used to discriminate between
phase equivalent $NN$ potentials  that differ essentially
off-shell like MEP and MP.
We suppose
that our result have a more general meaning showing that the detection of
hard photons accompanying $pp$ scattering at energies of $\sim$400~MeV
can be used to examine the off-shell properties of $NN$ interaction and to
test various models of $NN$ interaction, e.g., the ones
explicitly allowing for the quark degrees of freedom. Note, that the
experimental studies in this energy interval are planned for the nearest
future and some of them have been already started \cite{Nomachi}.

We study the coplanar $pp\to pp\gamma$ reaction
using the formalism of the
coordinate space representation described in detail elsewhere~\cite{YaF}.
The formalism makes it possible to avoid various approximations that are
used within the more conventional formalism of the momentum space
representation employed in
Refs.~\cite{Brown,Fearing,39,41,25,deJong,Eden,vonGeramb}.

Meson exchange currents are not allowed for in our calculations. The
contribution of the meson exchange currents to the $pp\to pp\gamma$ cross
section has been discussed in detail in Refs.~\cite{25,deJong,Eden}.
Meson exchange contribution is inessential at 280~MeV but increases with
energy. It is small for the emission of soft  photons, increases with the
photon energy, and decreases in the case of emission of the photons of
the maximal possible energy. At the same time, the off-shell differences
between the potential are most prominent in the latter case when the
energy of relative  motion of protons in the final state is small. This
corresponds to the kinematics when the angle between the protons in the
final state is small enough. So, the detection of protons at small angles
is favorable from the point of view of studying $NN$ interaction off-shell
while larger angles are more preferable for studying meson exchange
contributions.

%

%

Our results for ME Paris potential~\cite{Paris}, hard-core
Hamada--Johnston potential~\cite{H-J} and MP \cite{MoscowProgr} (see also
\cite{YaF} for improvement of MP higher partial waves)  are presented on
the figure. The predictions obtained with all above potentials
for the $pp\to pp\gamma$ differential cross sections at the incident proton
energy $\epsilon=280$~MeV are
very similar and describe well the experimental data of Ref.~\cite{expt3}.
The same conclusion
has been derived previously by Fearing~\cite{Fearing} who also involved
MP in the analysis of the $pp\to pp\gamma$ reaction cross section.


\begin{figure}[thb]
\centerline{
\parbox[t]{0.43\textwidth}{%
\psfig{figure=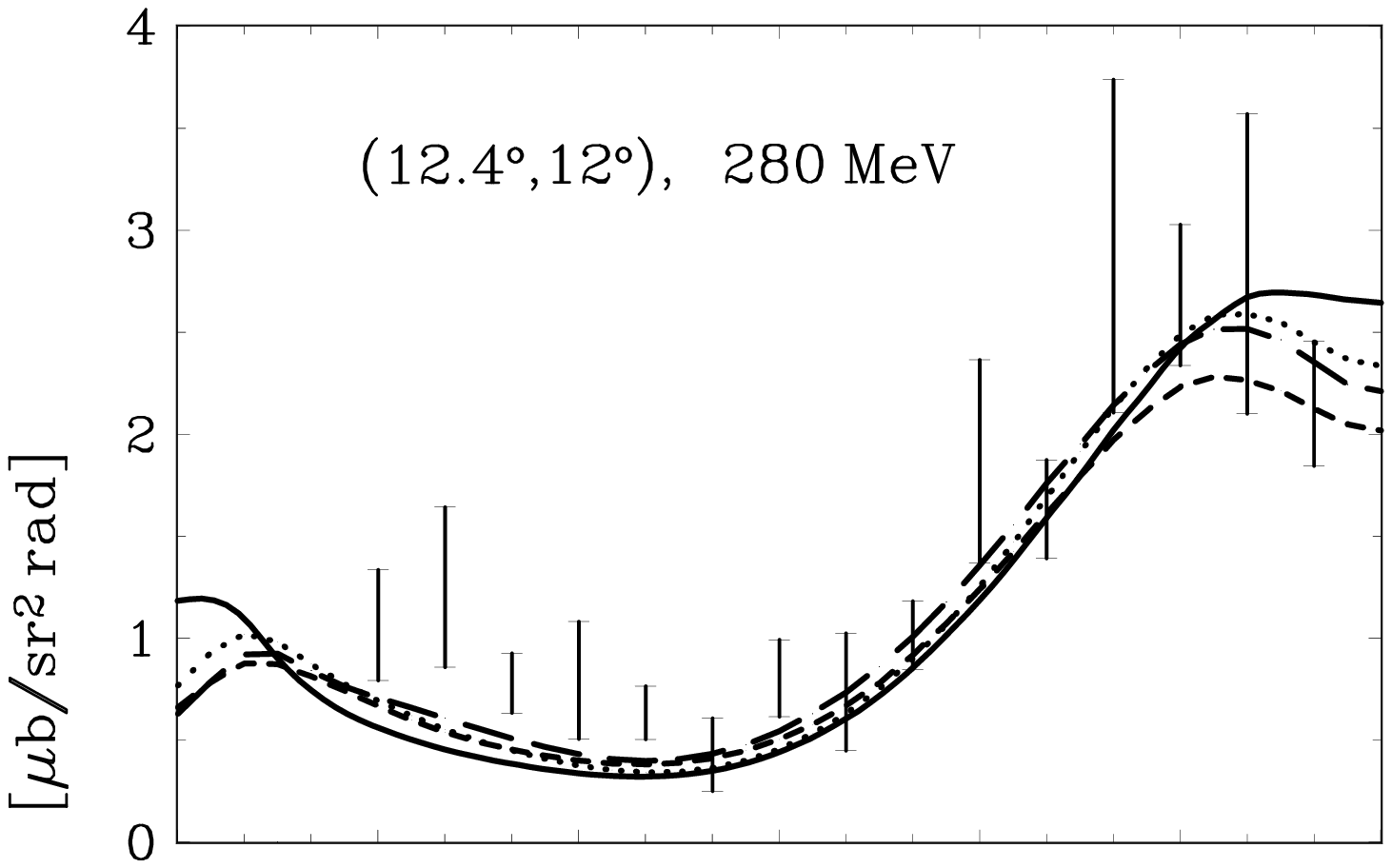,width=0.4\textwidth}%
\psfig{figure=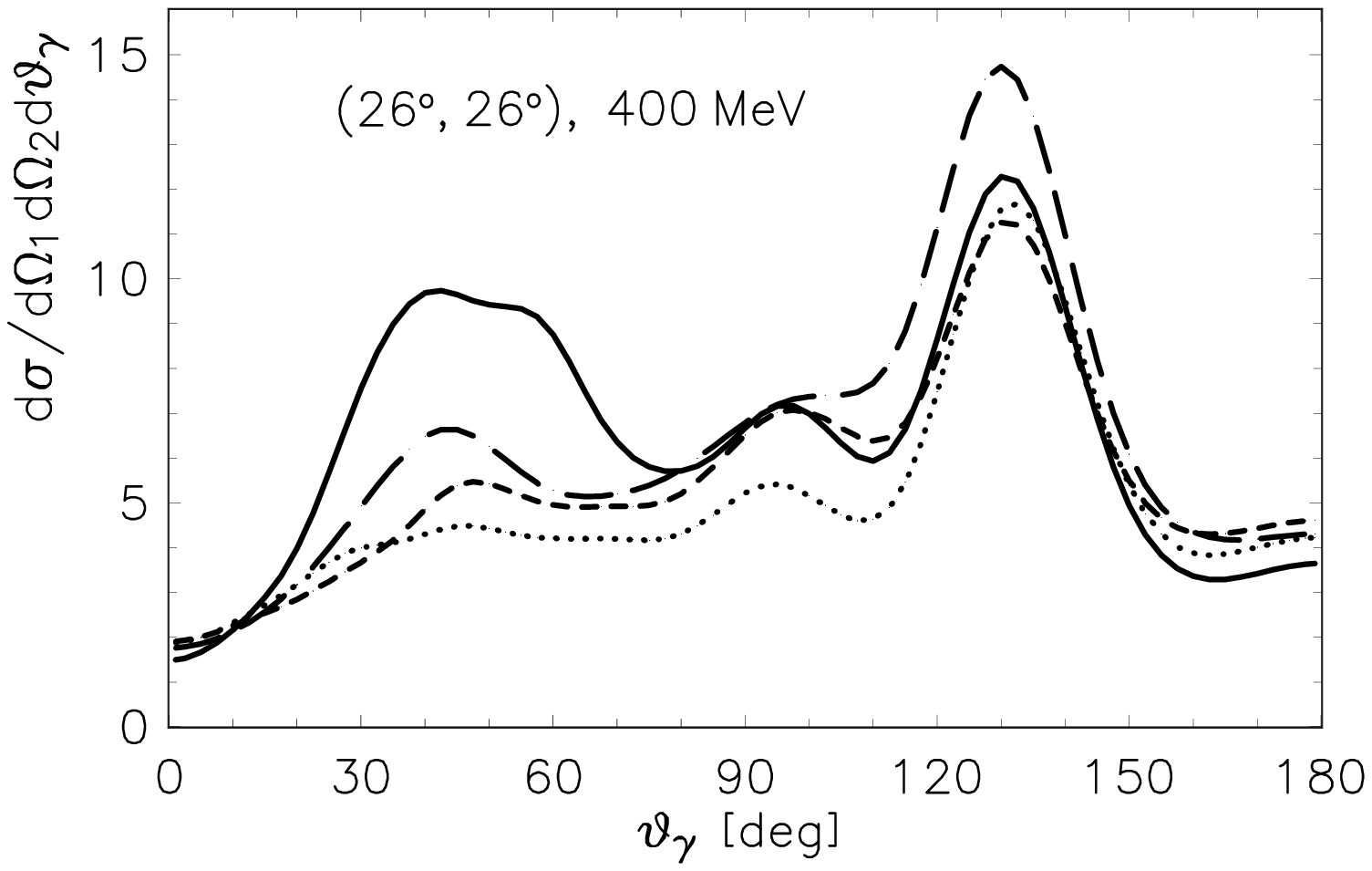,width=0.4\textwidth} }%
\hbox to 0.1\textwidth { }\parbox[t]{0.43\textwidth}{%
\psfig{figure=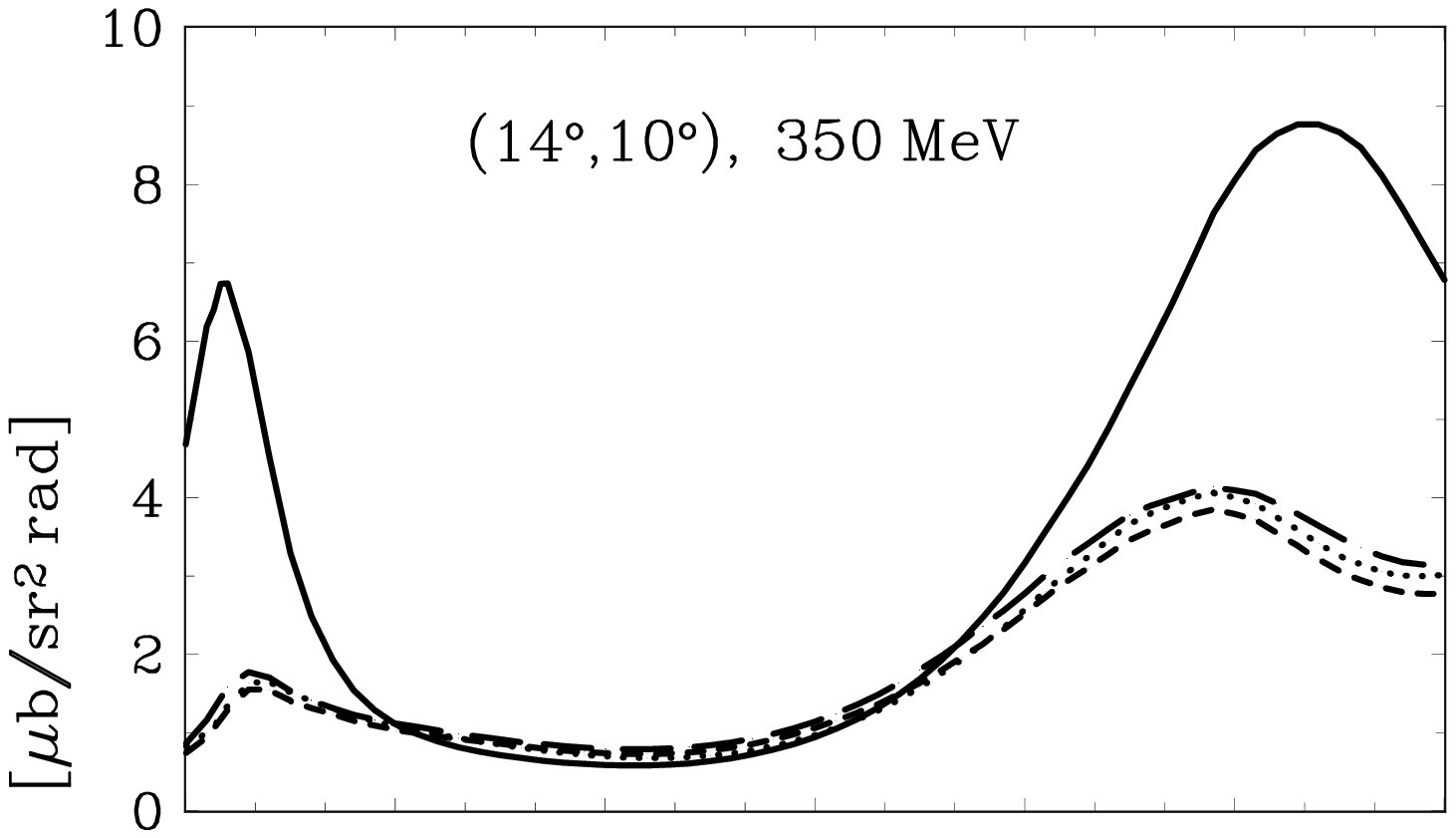,width=0.4\textwidth}%
\psfig{figure=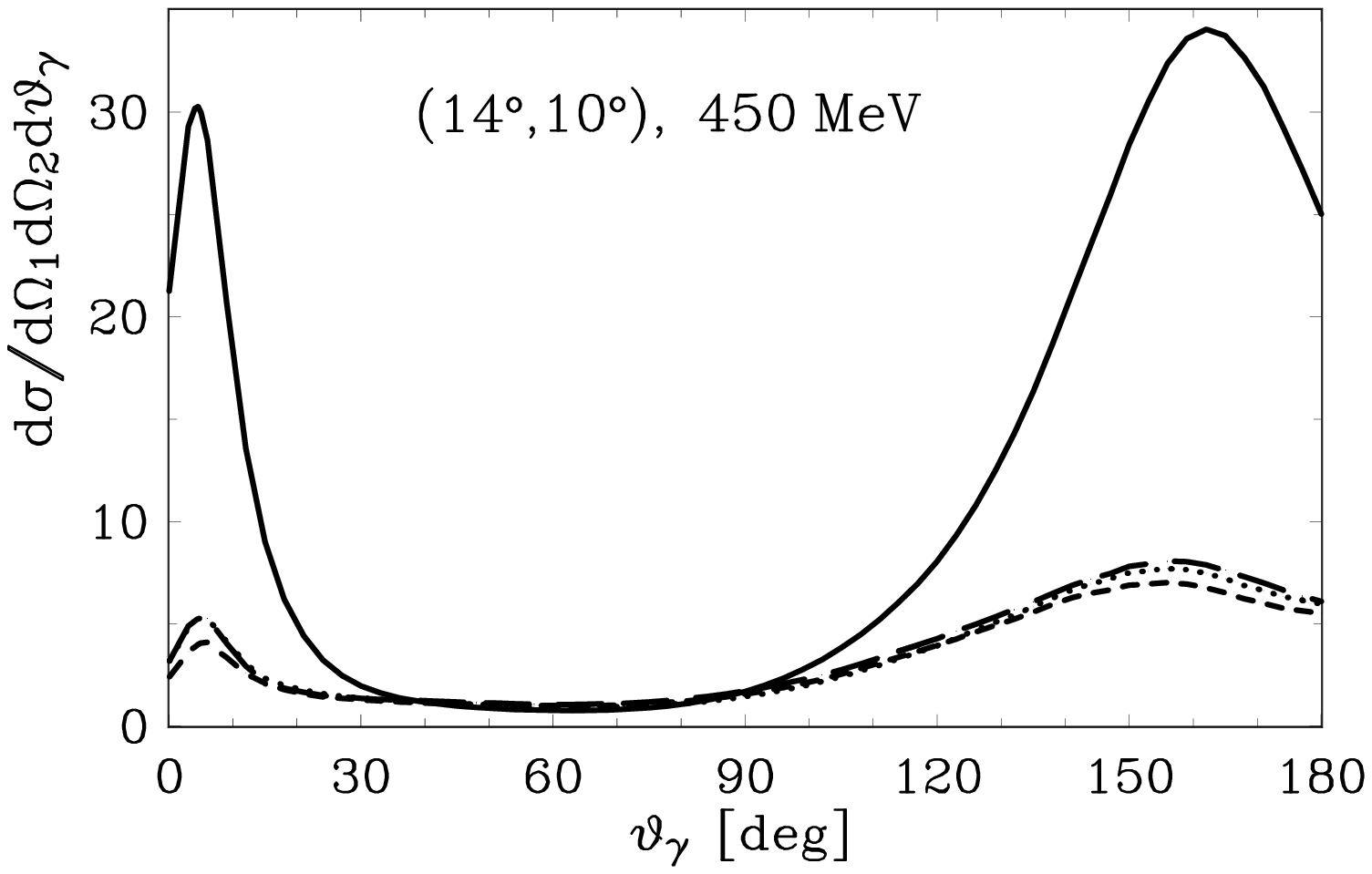,width=0.4\textwidth} }   }

\caption{ $pp\to pp\gamma$ cross section.  The angles of the final protons
are $\Theta_1$ and $\Theta_2$ and the beam energy $\epsilon$ are shown at each panel.
The results of
calculations with MP, super-symmetry partner of MP, Paris potential and
Hamada--Johnston potential are plotted by solid, dotted, long-dashed and
short-dashed lines, respectively.
Experimental data are taken from \protect\cite{expt3}.}
\label{280MeV}
\end{figure}

However, the situation changes drastically if the beam energy $\epsilon$
is increased up to 350 or 450~MeV. As is seen from the figure, the cross
section obtained with MP has well-pronounced peaks for the emission of
forward and backward photons that correspond to the maxima of the photon
energies in the C.M.S.; the respective maxima obtained with Paris and
Hamada--Johnston potentials are much less pronounced. The maximal values of
the cross section for MP and Paris or Hamada--Johnston potentials differ by
a factor of several times while the cross section for
$30^{\circ} < \Theta_{\gamma} < 90^{\circ}$ is the same for all potentials.
Therefore, to discriminate experimentally between MP and MEP in
the $pp\to pp\gamma$ reaction at the energy range of 350---450~MeV, one can
study the $\Theta_{\gamma}$-dependence only without absolute
normalization of the cross section. Note, that difference between MEP and MP
predictions is larger than the meson exchange contribution
even if it is calculated at larger energies $\epsilon=550$~MeV and larger proton angles (see
\cite{deJong}).

We have calculated also the MP super-symmetry partner \cite{SuSy}
that is exactly phase-equivalent to MP but supports the wave function
without the additional node like Paris potential. It is seen from the
figure that the results
obtained with the MP super-symmetry partner, Paris and Hamada--Johnston
potentials are nearly the same and differ essentially from the ones
obtained with MP. Thus, the
difference between MP and MEP predictions for $pp\to pp\gamma$ reaction
at energies of 350---450~MeV arise from the difference of the wave
functions at short range.  The difference is enhanced when the emitted
photon has the maximal possible C.M.S. energy for the given energy of the
proton beam. Therefore, the enhancement corresponds to the minimal C.M.S.
energy of the relative motion of the final protons. This kinematics
emphases the role of the $S$ and $P$ components in the $NN$ relative motion
in the final state which differ essentially at short range for MP and MEP.
The results of calculations for $\epsilon=400$~MeV correspond to the kinematical conditions
of the experiment started in Osaka \cite{Nomachi}. Unfortunately, the angle
between final protons in this experiment is large
($26^\circ +26^\circ =52^\circ$) and
the off-shell differences between the potential are less pronounced and are
comparable with the meson exchange contributions \cite{deJong}.

Summarizing, we have shown that the predictions for $pp\to pp\gamma$
reaction at the beam energies of 350---450~MeV obtained with the deep
attractive Moscow potential differ essentially from the ones obtained with
ME Paris and hard-core Hamada--Johnston potentials. So, the proton-proton
bremsstrahlung at energies of $\sim$400~MeV
can be used to study $NN$ interaction
off-shell and to discriminate between Moscow and ME potential models.


We are thankful to Y.~Mizuno, I.T.~Obukhovsky and V.N.~Pomerantsev for
discussions.
This research was supported in part by Russian Foundation for Basic
Research under the Grant No.~96--02--18072 and by Research
Program `Russian Universities'.


\end{document}